\documentclass[showpacs,preprint,floats,aps]{revtex4}

\usepackage{epsfig}
\usepackage{bm}

\begin{document}

\epsfclipon

\title{ Revisiting the Hugenholtz-Van Hove theorem in nuclear matter}
\author{ P. Czerski }
\affiliation{ Institute of Nuclear Physics, PL-31-342 Krak\'ow, Poland }
\author{ A. De Pace and A. Molinari }
\affiliation{
 Dipartimento di Fisica Teorica dell'Universit\`a di Torino and \\
 Istituto Nazionale di Fisica Nucleare, Sezione di Torino, \\
 via P.Giuria 1, I-10125 Torino, Italy 
}

\begin{abstract}
An assessment of the magnitude of the rearrangement contribution to
the Fermi energy and to the binding energy per particle is carried out in
symmetric nuclear matter by extending the $G$-matrix framework.
The restoration of the thermodynamic consistency or, equivalently, the 
fulfillment of the Hugenholtz--Van Hove theorem, is discussed.
\end{abstract}

\maketitle

\section{ Introduction}
\label{sec:intro}

The physics of nuclear matter beyond the Hartree-Fock (HF) mean field has been
dealt with a variety of techniques, such as the Bethe-Brueckner-Goldstone
hole-line expansion \cite{Bet71}, the variational approach based on correlated
functions \cite{Wir88} and the Green's function Monte Carlo method
\cite{Cep95}, exploiting realistic nucleon-nucleon interactions, like the Bonn
or the Urbana potentials, as input.

Basically, these studies focused on two observables: The binding energy per
particle and the saturation density, experimentally extracted from the 
semi-empirical mass formula and from the high energy nuclear electron
scattering experiments. In addition, also the compression modulus, whose value
can be inferred from the excitation energy of the breathing mode in nuclei, has
been quite extensively explored.

As it is well-known, all these investigations indicate that the theory fails 
to account for the data.
A remedy to this shortcoming is presently sought in the introduction of
three-body forces \cite{Sch86} and/or in a covariant treatment of the nuclear
many-body problem \cite{Ser86}. 

However, a successful theory of nuclear matter will be required not only to
account for the data, but to fulfill as well general theorems, in
particular the Hugenholtz-Van Hove (HvH) one \cite{Hug58}.
Indeed, a violation of the latter would signal an inconsistency of the theory
both at the global and at the single-particle level. 
 
In the following we shall focus on the latter issue, which has received 
comparatively less attention.
According to the HvH theorem, whose validity encompasses all the normal Fermi 
systems at zero temperature, at equilibrium the average energy per particle and
the Fermi energy should coincide. 

The HvH theorem holds because indeed a real (and not complex) energy can be 
assigned to the particles at the Fermi surface, as proved by Luttinger 
\cite{Lut61}.
Actually, the concept of single-particle energy is still approximately tenable 
in the proximity of the Fermi surface, but not away from the latter as
experimentally verified with exclusive inelastic electron scattering (e,e$'$p)
\cite{Bof96}.

Accordingly, in this paper we investigate 
\begin{itemize}
\item[a)] the contributions to the single particle energy (in particular at the
  Fermi surface) beyond the mean field,

\item[b)] how they affect the total binding energy per particle,

\item[c)] whether they help to fulfill the HvH theorem in nuclear matter.
\end{itemize}

These items have been considered in the past by various authors. In nuclear
matter starting from the old paper of Brueckner and Goldman \cite{Bru60a} till
the more recent work of Baldo {\em et al.\/} \cite{Bal90}. In finite nuclei,
where the issues are even more delicate, the theme has been addressed in
several investigations. To mention a few we recall those of Faessler {\em et
  al.\/} (see, for example, Ref.~\cite{Fae73}) and of Meldner and Shakin
\cite{Mel69}. A comprehensive review of the topic has been given by Hogdson
\cite{Hog94}. 

From the above studies it appears that, in a given theoretical framework, the
problem of fulfilling the HvH theorem is far from trivial, both in nuclear
matter and in finite nuclei. This recognition is the basic motivation for the
present work.

\section{ Formalism and $G$-matrix results}
\label{sec:forma}

We start from the G-matrix expression for the total energy of nuclear matter
at zero temperature
\begin{equation}
  \label{eq:En}
  E = \sum_{{\bm k}_1} \frac{k_1^2 }{ 2 m} n(k_1,k_F) + \frac{1 }{ 2} 
    \sum_{ \bm{k}_1 \bm{k}_2} <\bm{k}_1 \bm{k}_2 |G|\bm{k}_1 
    \bm{k}_2> n(k_1,k_F) n(k_2,k_F) , 
\end{equation} 
$k_F$ being the Fermi momentum and the G-matrix obeying the equation
\begin{eqnarray}
  \label{eq:Geq}
  &&<\bm{k}_1 \bm{k}_2|G|\bm{k}_1 \bm{k}_2> =
    <\bm{k}_1 \bm{k}_2|V|\bm{k}_1 \bm{k}_2> \nonumber \\
  && \qquad -\sum_{\bm{k}_3 \bm{k}_4} 
    \frac{<\bm{k}_1 \bm{k}_2|V|\bm{k}_3 \bm{k}_4> 
    (1 - n(k_3,k_F)) (1-n(k_4,k_F)) 
    <\bm{k}_3 \bm{k}_4|G|\bm{k}_1 \bm{k}_2>}{e_G (k_3,k_F) 
    + e_G (k_4,k_F) -  e_G (k_1,k_F) - e_G (k_2,k_F)} ,
\end{eqnarray} 
where the summation includes the spin-isospin degrees of freedom and the
matrix elements are meant to be antisymmetrized.
We then solve Eq.~(\ref{eq:Geq}) {\em self-consistently} with respect to the 
single-particle energies appearing in the denominator.
These, to be defined below, are continuous functions of the energy across the 
Fermi surface. For the bare interaction $V$ we use the Bonn potential
\cite{Mac89}. 

Concerning self-consistency, we recall that 
in Landau's theory of Fermi liquids the energy of a single quasi-particle 
$e(k,k_F)$ is obtained according to the prescription
\begin{equation}
  \label{eq:Landau}
  \frac{ \delta E }{ \delta n(k,k_F) } = e(k,k_F),
\end{equation}
where $n(k,k_F)$ gives the quasi-particle number in the $k$ state.
The distribution function $n(k,k_F)$ fixes the density of the system and
depends upon $k_F$, to be viewed as a parameter. 
When Eq.~(\ref{eq:Landau}) is implemented in the G-matrix framework (but
ignoring the dependence upon $n$ of the $G$-matrix elements), it yields
\begin{equation}
  \label{eq:eG}
  e_G (k,k_F) = \frac{ k^2 }{ 2 m } + \sum_{\bm{k}_2}  
    <\bm{k} \bm{k}_2|G|\bm{k} \bm{k}_2> n(k_2,k_F),
\end{equation}
which we view as the energy of a particle of given momentum $k$.

In accord with the theory of Brueckner \cite{Bru58}, which represent the
leading term of the hole-line expansion and exactly incorporates the
two-particle correlations, we solve then Eq.~(\ref{eq:Geq}) seeking for 
self-consistency on the basis of Eq.~(\ref{eq:eG}).
From a diagrammatic point of view this amounts to dress the particles and the
{\em two-hole} lines of a ladder diagram with a {\em first order} self-energy
containing a $G$-matrix interaction.

However, it is known from a long time that, when computed at $k=k_F$, 
Eq.~(\ref{eq:eG}) violates the HvH theorem, in contrast with the HF theory 
(see the Appendix). This is indeed seen to occur in
Fig.~\ref{fig:fig_EG}, where we display, as a function of
$\rho=2k_F^3/3\pi^2$, the results 
of our calculation of the binding energy per particle (\ref{eq:En}) and of the
Fermi energy, as obtained from (\ref{eq:eG}), carried out with the distribution
$n(k,k_F)=\theta(k_F-k)$.
It appears from the figure that our self-consistent $G$-matrix yields a good 
binding energy per particle (-16.1 MeV), but at the wrong density 
($\rho=0.25$ fm$^{-3}$, i.~e. $k_F=1.55$ fm$^{-1}$); these results are
close to those recently obtained in Ref.~\cite{Son98}, also in the
Brueckner-Hartree-Fock (BHF) scheme, with the $v14$ Argonne potential and with
a continuous auxiliary potential. In addition, a most substantial violation of
the HvH theorem (of 17.2 MeV, the Fermi energy being $\epsilon_F=-33.3$ MeV) is
seen to occur, {\em in quantitative} accord with the findings of 
Ref.~\cite{Bal90}.

Note that the above quoted $\epsilon_F$ corresponds to a BHF potential energy
of about 80 MeV: Hence, the rearrangement corresponds to a correction of
$\cong20\%$ of the latter.

\begin{figure}[t]
\begin{center}
\mbox{\epsfig{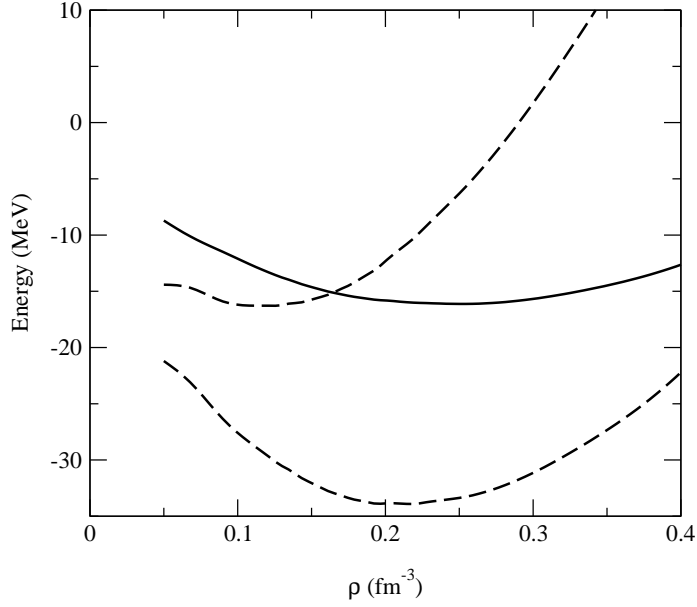}}
\caption{Symmetric nuclear matter binding energy per particle (solid) and Fermi
  energy (dashed) as a function of the density. The lower and upper dashed
  lines refer to Fermi energies without and with rearrangement contribution,
  respectively, whereas the solid line has no rearrangement. 
}
\label{fig:fig_EG}
\end{center}
\end{figure}

Thus, in the spirit of Landau's theory, we carry out more thoroughly the
variation (\ref{eq:Landau}) of the energy versus the distribution $n(k,k_F)$ 
writing
\begin{equation}
  e(k,k_F) = e_G (k,k_F) + \frac{1 }{ 2} \sum_{\bm{k}_1 \bm{k}_2} 
    n(k_1,k_F) n(k_2,k_F) \frac{ \delta  }{ \delta n(k,k_F) }
    <\bm{k}_1 \bm{k}_2|G|\bm{k}_1 \bm{k}_2>, 
\end{equation}
the last term on the right hand side being often referred to as the {\em 
rearrangement} contribution. This in principle should be computed by making
the variation on the right hand side of Eq.~(\ref{eq:Geq}) with respect to
$n(k_3,k_F)$ and $n(k_4,k_F)$ and also to the $n(k,k_F)$ entering into the
single particle energies of the denominator.
These variations represent corrections to the Pauli operator and to the
energies of the initial and intermediate states, respectively, resulting from
the removal of one particle.

For a discussion of the associated diagrams and for the convergence of 
the hole-line expansion in general we refer the reader to
Refs.~\cite{Son98,Bal90,Bro71,Boz01}, where the impact of the hole-hole
ladder diagrams and of the off-shell $T$-matrix in connection with the HvH 
theorem is explored.

Here we aim instead to compute the rearrangement contribution to the single 
particle energy at (or in the proximity of) the Fermi surface by using the 
following procedure in carrying out the functional derivative, that is
\begin{equation}
  \label{eq:erear}
  e(k\approx k_F,k_F) = e_G (k,k_F) + 
    \frac{1 }{ 2} \sum_{\bm{k}_1 \bm{k}_2} 
    n(k_1,k_F) n(k_2,k_F) \frac{ \partial }{ \partial k_F }
    <\bm{k}_1 \bm{k}_2|G|\bm{k}_1 \bm{k}_2>  
    \frac{ \partial  k_F }{ \partial \rho } \frac{\delta\rho}{\delta n(k,k_F)},
\end{equation}
where $\rho\equiv\rho(k_F)$ is the density of the system. Of course, if the
(large) volume $V$ enclosing the nuclear matter is kept fixed, then 
\begin{equation}
  \label{eq:deltarho}
  \frac {\delta \rho }{ \delta n(k,k_F) } =  \frac{1 }{ V}
    \frac{\delta}{\delta n(k,k_F)} \sum_{\bm{k}_1} n(k_1,k_F)
    = \frac{1 }{ V} 4 \frac{V }{ (2\pi)^3 }\int d\bm{k}_1 
    \frac{\delta n(k_1,k_F) }{ \delta n(k,k_F)} = \frac{1 }{ V}.
\end{equation}
In the Appendix we shall show that the above prescription for the functional 
derivative of the energy with respect to the distribution, worked out at 
constant volume, just yields the Fermi energy, whereas, when the number of 
particles is kept fixed, is proportional to the pressure.
The latter, of course, should vanish for a system at equilibrium.

To proceed further we now define
\begin{equation}
  \label{eq:A}
  A(k_F) = \frac{1}{2N}\sum_{\bm{k}_1\bm{k}_2}n(k_1,k_F)n(k_2,k_F)
    \frac{\partial}{\partial k_F}
    \langle\bm{k}_1,\bm{k}_2|G|\bm{k}_1,\bm{k}_2\rangle,
\end{equation}
which, together with Eq.~(\ref{eq:deltarho}), allows us to recast
(\ref{eq:erear}) as follows:
\begin{eqnarray}
  \label{eq:edlog}
  e(k\approx k_F,k_F) &=& e_G (k,k_F) + 
    \rho\frac{\partial k_F}{\partial\rho}A(k_F) \nonumber \\
  &=& e_G (k,k_F) + \left(\frac{\partial}{\partial k_F}\ln\rho\right)^{-1}
    A(k_F).
\end{eqnarray}
The above equation, where the $G$-matrix and the rearrangement contributions
to the single particle (or, better, quasi-particle) energy are neatly
separated, yields the correct Fermi energy $e(k_F,k_F)\equiv\epsilon_F$, being
rigorously valid at the Fermi surface. 

We should now compute the impact of the rearrangement on the binding energy per
particle. In this connection, we are aware that the BHF formula
\begin{equation}
  \label{eq:EG}
  \frac{E_G}{N}=\frac{1}{2N}\sum_{\bm{k}}n(k,k_F)\left[
    \frac{k^2}{2m}+e_G(k,k_F)\right],
\end{equation}
--- linking the total and the single particle energies, --- no longer holds 
when the rearrangement is included. Furthermore, calculations of the nucleon 
momentum distribution including correlations among nucleons beyond the BHF
ones indicate that the corrections to the latter tend to be more pronounced 
near the Fermi surface \cite{Die90}, where the Landau theory applies.
Hence, inspired by the Landau theory, 
we heuristically assume Eq.~(\ref{eq:edlog}) to be applicable as well in
the proximity of the Fermi surface, where the concept of quasi-particle is
tenable. In the conclusions the foundation and the limits of this assumption
will be addressed. Here we write, as an extension of (\ref{eq:EG}),
\begin{eqnarray}
  \label{eq:EQP}
  \frac{E}{N} &=& \frac{E_G}{N} + A(k_F) 
    \left(\frac{\partial}{\partial k_F}\ln\rho\right)^{-1}\frac{1}{2N}
    \sum_{k>k_{\text{QP}}} n(k,k_F) \nonumber \\
  &=& \frac{E_G}{N} + A(k_F) 
    \left(\frac{\partial}{\partial k_F}\ln\rho\right)^{-1}\frac{1}{\pi^2\rho}
    \int_{k_{\text{QP}}}^{k_F}dk\,k^2 \\
  &=& \frac{E_G}{N} + A(k_F) 
    \left(\frac{\partial}{\partial k_F}\ln\rho\right)^{-1}\frac{1}{2}
    \left[1-\left(\frac{k_{\text{QP}}}{k_F}\right)^3\right]. \nonumber
\end{eqnarray}
In the above the summation is meant to be carried out in a restricted region
near the Fermi surface, namely in the range of momenta 
$k_{\text{QP}}\le k\le k_F$. In the derivation of (\ref{eq:EQP}), the
$\theta$-function distribution is used to be consistent with the $G$-matrix
calculation. Obviously the momentum $k_{\text{QP}}$ should be viewed as a
parameter: If it will turn out to be close to $k_F$, then the omission of the
$k$-dependence in the rearrangement contribution to the single-particle energy
(\ref{eq:edlog}) should be expected to have not too serious consequences.

Thus, the expression (\ref{eq:EQP}), beyond the standard $G$-matrix 
contribution (namely $E_G/N$), explicitly embodies the rearrangement one as 
well, but the latter, --- and this is important, --- is reduced by the factor 
$[1-(k_{\text{QP}}/k_F)^3]/2$ with respect to Eq.~(\ref{eq:edlog}).

\begin{figure}[t]
\begin{center}
\mbox{\epsfig{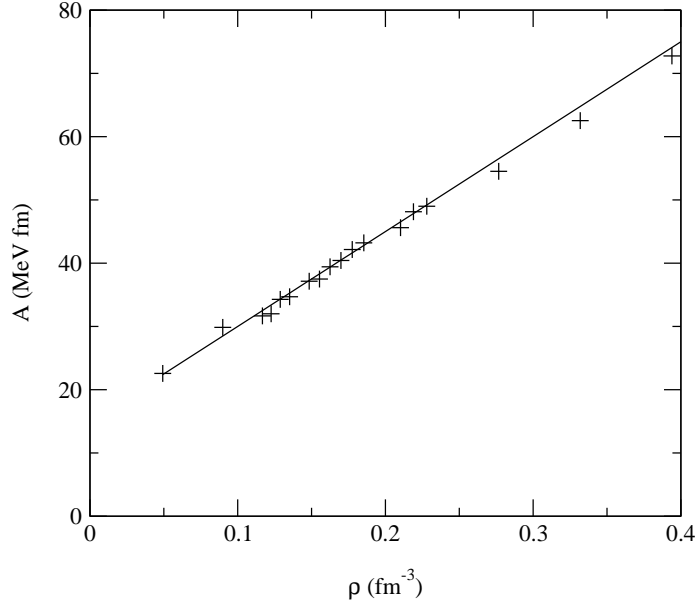}}
\caption{The quantity $A(k_F)$, defined in the formula (\protect\ref{eq:A}) of
  the text versus the density $\rho=2k_F^3/2\pi^2$. The crosses show the
  calculated values, whereas the solid line corresponds to the linear fit
  $A=150\rho+15$ MeV fm. We do not display $A(k_F)$ for $\rho<0.05$ fm$^{-3}$ 
  because here its numerical evaluation becomes quite inaccurate.
}
\label{fig:fig_a}
\end{center}
\end{figure}

\section{ Results for the rearrangement energy}
\label{sec:resul}

In order to compute the rearrangement contribution, the quantity $A(k_F)$ is
needed: It has been numerically evaluated and it is displayed as a function of
$\rho$ in Fig.~\ref{fig:fig_a}. It appears that $A(k_F)$ is always positive,
growing linearly with $\rho$, and substantial. Actually, it turns out to be
well approximated by the expression $A(k_F)=(150\rho+15)$ MeV fm.

\begin{figure}[t]
\begin{center}
\mbox{\epsfig{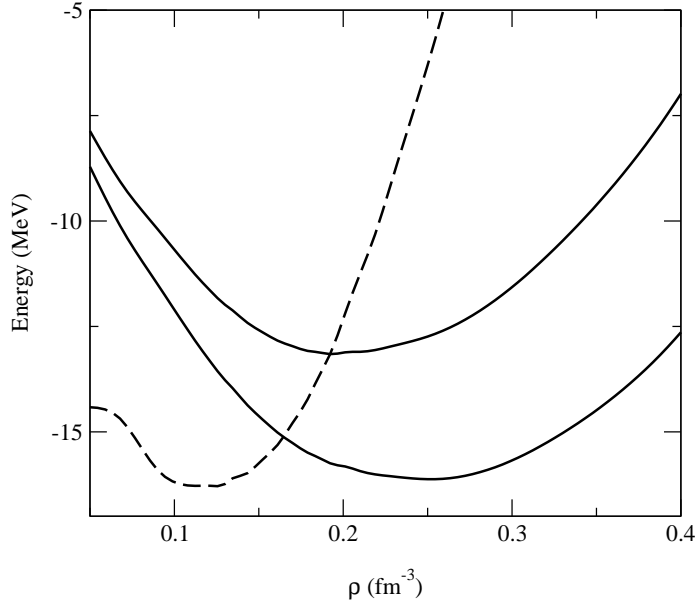}}
\caption{Symmetric nuclear matter binding energy per particle (solid) and Fermi
  energy (dashed) as a function of the density. The lower and upper solid
  lines refer to binding energies without and with rearrangement contribution,
  respectively, whereas the dashed line includes rearrangement.
}
\label{fig:fig_EQP}
\end{center}
\end{figure}

Since the above findings on $A(k_F)$ have been obtained with a
$\theta$-function distribution, for consistency we evaluate
$[\partial(\ln\rho)/\partial k_F]^{-1}$ with the $\theta$-function as well.
An elementary calculation yields 
\begin{equation}
  \label{eq:dlogtheta}
  \left(\frac{\partial}{\partial k_F}\ln\rho\right)^{-1} = \frac{k_F}{3}.
\end{equation}
Hence, from Eq.~(\ref{eq:edlog}) it follows that the rearrangement contribution
to the Fermi energy grows as $k_F^4$, in accord with Ref.~\cite{Tho58}, where 
it was conjectured to grow as $k_F^5$ and found to vary as $k_F^4$ in a
schematic estimate.  

In Fig.~\ref{fig:fig_EG}, we display the Fermi energy including the 
rearrangement, obtained using (\ref{eq:dlogtheta}), versus the density 
$\rho=2k_F^3/3\pi^2$.

We note: 

i) The occurrence, in correspondence of the minimum of the binding
energy, of a {\em positive} rearrangement contribution of about 25 MeV, 
larger than the value of 17 MeV obtained in Ref.~\cite{Bal90} in a perturbative
scheme. Since in Ref.~\cite{Bal90} only the leading diagrams, beyond the BHF
ones, were kept, one could view the difference as an estimate of the
contribution of the higher order terms in the hole-line expansion. However,
this estimate should not be taken at face value, since in \cite{Bal90} the
Paris potential was employed, whereas we use the Bonn one; 

ii)notably, our
rearrangement contribution stays remarkably constant to the left of our
saturation density $\rho\approx0.25$ fm$^{-3}$ (say, in the range
$0.20\le\rho\le0.25$ fm$^{-3}$), in accord with the old finding of Thouless
\cite{Tho58}; 

iii) while in Ref.~\cite{Bal90} the obtained rearrangement
contribution was just enough to restore the thermodynamic consistency, in our
case we ``overcure'' the HvH violation: Indeed, we predict the equality
between the Fermi energy and the binding energy to occur at $\rho\approx0.17$
fm$^{-3}$, i.~e. not at equilibrium. However, in \cite{Bal90} no change 
was assumed to occur in the ground state energy due to rearrangement. 

Here, as discussed above, we schematically estimate the magnitude of this 
change by resorting to formula (\ref{eq:EQP}) and searching for a value of the
momentum $k_{\text{QP}}$ such to restore the validity of the HvH theorem. 
This turns out to be fulfilled when $k_{\text{QP}}=0.09 k_F$, as it appears
from Fig.~\ref{fig:fig_EQP}, where we display, in correspondence to this value,
the binding and Fermi energies, both including rearrangement, versus the
density. 

We see in the figure that the minimum of the binding energy, while reduced, as
expected, to a value (-13.15 MeV) higher than the BHF prediction, occurs at a
saturation density $\rho_{\text{min}}\cong0.19$ fm$^{-3}$ ($k_F=1.41$
fm$^{-1}$), in closer contact with the experimental value
$\rho_{\text{exp}}=0.17$ fm$^{-3}$.

Concerning the rearrangement effect, at $k_F=1.41$ fm$^{-1}$ we obtain a
positive contribution of $\approx20$ MeV for the Fermi energy and of
$\approx1.8$ MeV for the binding energy per particle.
These values might be compared with those obtained long ago by Brueckner {\em
  et al.\/} \cite{Bru60b}, namely about 12 and 1.5 MeV for the Fermi and mean
energies, respectively, and with the recent ones of Ref.~\cite{Boz01}, namely
about 12 and 1 MeV, respectively. It should however be remarked that these
values have been obtained with simple, as compared to the Bonn potential,
interactions. 

Finally, the compression modulus predicted in our framework, via a polynomial
interpolation of the binding energy, turns out to be $\approx150$ MeV, 
significantly larger than the one obtained in a pure BHF scheme 
($\approx120$ MeV), as expected owing to the rapid $k_F$ dependence of the
rearrangement term, but still somewhat lower than the experimental value.

\section{ Conclusions }
\label{sec:concl}

In these concluding remarks we seek for some justification of our empirical
procedure in computing the rearrangement contribution to the binding energy per
particle. 

We start by recalling the link (see (\ref{eq:EG})) between the single-particle
and mean energies, which holds only for strict mean filed theories, like HF and
BHF. Should these schemes be valid, then, as it is well-known \cite{Hog94}, the
separation and the single-particle energies would coincide, as it is (nearly)
true in atoms (Koopman's theorem, see Ref.~\cite{Hog94}).

In nuclei, of course, HF (which respects HvH) is not applicable and BHF, as
shown by many calculations, while not unrealistic, fails to fulfill the HvH
theorem. Indeed, the amount of the failure measures the impossibility of
describing the system in terms of independent constituents.
Actually, beyond the mean field framework the only remaining link between the
mean and single-particle energies is the one expressed by the HvH theorem
itself. 

Our simple approach is based on the premise that the contribution to the
system's mean energy arising from the proximity of the Fermi surface can still
be simply related to the energy of individual entities.
We identify the latter with Landau's quasi-particles and not with the BHF
particles, since, as previously mentioned, the BHF framework fails to account 
for the appreciable depletion of the single-particle orbits induced by the 
strong short-range repulsion among the nucleons, a depletion occurring mainly,
although not only, near the Fermi surface.
Indeed, it is established that the repulsion much affects both the
momentum distribution and the rearrangement energy \cite{Tho58}.

Accordingly, we use the quasi-particle approximation for the propagator
$G(\bm{k},\omega)$ in the expression \cite{Fet71}
\begin{equation}
  \frac{E}{N} = \frac{1}{\rho} \lim_{\eta\to0^+}
    \int\frac{d\bm{k}}{(2\pi)^3} \int_{-\infty}^{\infty}
    \frac{d\omega}{2\pi i} e^{i\omega\eta}\left[\frac{k^2}{2m}+\frac{1}{2}
    \Sigma^*(\bm{k},\omega)\right]\text{Tr}\, G(\bm{k},\omega),
\end{equation}
which yields the binding energy per particle.
The quasi-particle approximation for $G(\bm{k},\omega)$ is best grasped by
starting from the canonical spectral representation,
\begin{equation}
  G(\bm{k},\omega) = \int_{\epsilon_F}^{\infty} d\omega'
    \frac{S_{\text{p}}(\bm{k},\omega')}{\omega-\omega'+i\eta} +
    \int_{-\infty}^{\epsilon_F} d\omega'
    \frac{S_{\text{h}}(\bm{k},\omega')}{\omega-\omega'-i\eta}.
\end{equation}
It amounts to set for the hole spectral function the expression
\begin{equation}
  S_{\text{h}}^{\text{QP}}(\bm{k},\omega) = \frac{1}{\pi}
    \frac{Z^2(k)|W(k)|}{[\omega-E(k)]^2+[Z(k)W(k)]^2},
\end{equation}
where $W(k)$ is the imaginary part of the quasi-particle self-energy and $Z(k)$
the so-called quasi-particle strength.

In our crude and empirical model, to avoid the introduction of further 
parameters beyond
$k_{\text{QP}}$, we have set $Z(k)=1$ and $W(k)=0$. Of course, the former
should be lower than one, but not too much, in order not to spoil the concept
of quasi-particle, and the latter should be small for the same reason.
Thus our quasi-particle propagator differs from the BHF one only in the 
location of the pole, which is moved by the rearrangement contribution.

Finally, we find it gratifying that the only parameter of our approach,namely
$k_{\text{QP}}$, confines the quasi-particle existence to a quite narrow domain
close to the Fermi surface.
Also satisfying is our result that the rearrangement affects the Fermi energy
an order of magnitude more than the mean energy, a finding on which a general
consensus exists.

\appendix*

\section{}

We comment here on the formula
\begin{equation}
  \label{eq:deltaE}
  \frac{\delta E[n]}{\delta n(k,k_F)}\Big|_{k\cong k_F} = 
    \frac{\partial E}{\partial k_F}
    \frac{\partial k_F}{\partial\rho} \frac{\delta\rho}{\delta n(k,k_F)},
\end{equation}
where
\begin{equation}
  \frac{\delta\rho}{\delta n(k,k_F)} = 
    \frac{\delta}{\delta n(k,k_F)} \left[\frac{1}{V}\sum_{\bm{k}_1}
    n(k_1,k_F) \right],
\end{equation}
for the functional derivative of the energy with respect to the distribution
function. The sum over spin and isospin is understood.

We first consider the volume $V$ constant. In this case
\begin{equation}
  \frac{\delta\rho}{\delta n(k,k_F)} = \frac{1}{V}\sum_{\bm{k}_1}
    \frac{\delta n(k_1,k_F)}{\delta n(k,k_F)} = \frac{1}{V}.
\end{equation}
Hence
\begin{equation}
  \label{eq:deltaEV}
  \frac{\delta E[n]}{\delta n(k,k_F)}\Big|_{k\cong k_F} = 
    \frac{\partial E}{\partial k_F}
    \frac{\partial k_F}{\partial (N/V)} \frac{1}{V} =
    \frac{\partial E}{\partial k_F} \frac{\partial k_F}{\partial N} =
    \frac{\partial E}{\partial N} = \mu = \epsilon_F.
\end{equation}
In the case of a free Fermi gas, (\ref{eq:deltaEV}) yields indeed
\begin{equation}
  \frac{\delta E[n]}{\delta n(k,k_F)}\Big|_{k\cong k_F} = 
    \frac{\partial E}{\partial k_F}\frac{\pi^2}{2k_F^2}\frac{1}{V} = 
    \frac{\pi^2}{2k_F^2}\frac{\partial}{\partial k_F}\left(\frac{N}{V}
    \frac{3}{5}\frac{k_F^2}{2m}\right) = 
    \frac{k_F^2}{2m} \equiv \epsilon_F^{\text{FG}}.
\end{equation}
On the other hand, when $N$ is constant (but $V$ varies)
\begin{equation}
  \delta N = \sum_{\bm{k}}\frac{\delta N}{\delta n(k,k_F)} 
    \delta n(k,k_F) =
    \sum_{\bm{k}}\delta n(k,k_F)=0,
\end{equation}
since
\begin{equation}
  \frac{\delta N}{\delta n(k,k_F)} = 
    \frac{\delta}{\delta n(k,k_F)} \sum_{\bm{k}_1}n(k_1,k_F)=1.
\end{equation}
Thus, the vanishing of $\delta N$ does not imply the vanishing of 
$\delta N/\delta n(k,k_F)$. Hence
\begin{equation}
  \label{eq:deltaEN1}
  \frac{\delta E[n]}{\delta n(k,k_F)}\Big|_{k\cong k_F} = 
    \frac{\partial E}{\partial k_F}
    \frac{\partial k_F}{\partial\rho}\frac{1}{V} - 
    \frac{\partial E}{\partial k_F}\frac{\partial k_F}{\partial\rho}
    \frac{\rho}{V}\frac{\delta V}{\delta n(k,k_F)},
\end{equation}
the second term on the right hand side actually not contributing because $V$ 
and $n(k,k_F)$ are varying independently. Accordingly
\begin{equation}
  \label{eq:deltaEN2}
  \frac{\delta E[n]}{\delta n(k,k_F)}\Big|_{k\cong k_F} = 
    \frac{\partial E}{\partial k_F}
    \frac{1}{N}\frac{\partial k_F}{\partial (1/V)}\frac{1}{V} =
    -\frac{1}{\rho}\frac{\partial E}{\partial V} = \frac{P}{\rho},
\end{equation}
which vanishes at equilibrium because so does the pressure $P$.

Curiously, from (\ref{eq:deltaEN2}) for a perfect Fermi gas it follows 
\begin{equation}
  \frac{\delta E[n]}{\delta n(k,k_F)}\Big|_{k\cong k_F} = 
    \frac{2}{5}\epsilon_F = \epsilon_F-\frac{3}{5}\epsilon_F,
\end{equation}
showing that a non confined Fermi gas satisfies the HvH theorem,
i.~e. it reaches equilibrium, at zero density.

On the other hand, for a translationally invariant Fermi system, with a generic
interaction $V(q)$, one has from (\ref{eq:deltaEN2})
\begin{equation}
  \label{eq:deltaENint}
  \frac{\delta E[n]}{\delta n(k,k_F)}\Big|_{k\cong k_F} = 
    \frac{k_F}{3}\frac{\partial(E/N)}{\partial k_F} = 0.
\end{equation}
In the HF approximation \cite{Fet71}
\begin{eqnarray}
  \frac{E}{N} &=& \frac{3}{5}\epsilon_F^{\text{FG}} + \frac{1}{2}\rho V(0) -
    \int\frac{d\bm{k}}{(2\pi)^3}\theta(k_F-k)
    \int\frac{d\bm{k}^\prime}{(2\pi)^3}\theta(k_F-k^\prime)
    V(|\bm{k}-\bm{k}^\prime|) \nonumber \\
  &=& \frac{3}{5}\epsilon_F^{\text{FG}} + \frac{1}{2}\rho V(0) -
    \frac{1}{(2\pi)^2}\int_0^{2k_F}dq\, q^2 V(q) \left[
    1-\frac{3}{2}\frac{q}{2k_F}+\frac{1}{2}\left(\frac{q}{2k_F}\right)^3\right]
\end{eqnarray}
and it is an easy matter to carry out the derivative in (\ref{eq:deltaENint}),
getting 
\begin{equation}
  \label{eq:derEN}
  \frac{\partial(E/N)}{\partial k_F} = -\frac{E}{N} + \epsilon_F^{\text{FG}}
    + \rho V(0) - \frac{1}{(2\pi)^2}\int_0^{2k_F}dq\, q^2 V(q) 
    \left(1-\frac{q}{2k_F}\right).
\end{equation}
Now, in HF the Fermi energy reads
\begin{eqnarray}
  \epsilon_F^{\text{HF}} &=& \epsilon_F^{\text{FG}} + \rho V(0) -
    \int\frac{d\bm{k}^\prime}{(2\pi)^3}\theta(k_F-k^\prime)
    V(|\bm{k}_F-\bm{k}^\prime|) \nonumber \\
  &=& \epsilon_F^{\text{FG}} + \rho V(0) -
    \frac{1}{(2\pi)^2}\int_0^{2k_F}dq\, q^2 V(q) \left(1-\frac{q}{2k_F}\right).
\end{eqnarray}
Hence, by comparing with (\ref{eq:derEN}), one sees that the HvH theorem is
fulfilled in the HF approximation no matter what the interaction is, providing
the latter is independent of the density.

\end{document}